\documentclass[10pt,twocolumn,letterpaper]{IEEEtran}
\usepackage{graphicx}
\usepackage{epsfig}
\usepackage{latexsym}
\usepackage{amsfonts}
\usepackage{here}
\usepackage{rawfonts}
\usepackage[latin1]{inputenc}
\usepackage[T1]{fontenc}
\usepackage{calc}
\usepackage{url}
\usepackage{enumerate}
\usepackage{color}
\usepackage{amssymb}
\usepackage{bm}
\usepackage{upref}
\usepackage{times}
\usepackage{amsmath}
\usepackage{cite}

\newcommand{\bs}{\mathbf}
\newcommand{\ms}{\;\;}

\newcommand{\mycaption}[1]{\vspace*{-1.6em}\caption{#1}\vspace*{-1.0em}}

\newtheorem{theorem}{{\bf Theorem}}

\newcommand{\qed}{\nobreak \ifvmode \relax \else
  \ifdim\lastskip<1.5em \hskip-\lastskip
  \hskip1.5em plus0em minus0.5em \fi \nobreak
  \vrule height0.75em width0.5em depth0.25em\fi}

\newcounter{step}
\newlength{\totlinewidth}
  {\end{list}%
  \rule{\linewidth}{1pt}}
\newcounter{substep}

  {\end{list}}

\newlength{\aligntop}
\newlength{\alignbot}
\makeatletter
\renewenvironment{align}{%
  \vspace{\aligntop}
  \start@align\@ne\st@rredfalse\m@ne
}{%
  \math@cr \black@\totwidth@
  \egroup
  \ifingather@
    \restorealignstate@
    \egroup
    \nonumber
    \ifnum0=`{\fi\iffalse}\fi
  \else
    $$%
  \fi
  \ignorespacesafterend%
  \vspace{\alignbot}\par\noindent
}
\makeatother

\IEEEoverridecommandlockouts

\author{Manav R. Bhatnagar,~\IEEEmembership{Member,~IEEE} 
\vspace*{-2.6em}%
  \thanks{\textbf{Corresponding author:} Manav R. Bhatnagar is with
    Department of Electrical Engineering, Indian Institute of Technology Delhi, Hauz Khas, IN-110016 New Delhi, India,
    email: \protect\url{manav@ee.iitd.ac.in}.}% 
}\date{}
\renewcommand{\baselinestretch}{1.0}
\title{Performance Analysis of Decode-and-Forward Relaying in Gamma-Gamma Fading Channels\vspace*{-0.25em}}

\begin{document}

\maketitle
\vspace*{-1em}
\begin{abstract}
Decode-and-forward (DF) cooperative communication based on free space optical (FSO) links is studied in this letter. We analyze performance of the DF protocol in the FSO links following the Gamma-Gamma distribution. The cumulative distribution function (CDF) and probability density function (PDF) of a random variable containing mixture of the Gamma-Gamma and Gaussian random variables is derived. By using the derived CDF and PDF, average bit error rate of the DF relaying is obtained.   
\end{abstract}\vspace*{-0.5em}
{\bf Index Terms:} Bit error rate, decode-and-forward relaying, free space optical links, Gamma-Gamma fading.
\vspace*{-1.0em}  
\section{Introduction}\vspace*{-0.25em}
Free space optical communication has recently drawn significant attention because of very large bandwidth,
low implementation cost, and excellent security~\cite{kedar04,andre01}.  
The intensity modulation and direct detection (IM/DD) with on/off keying (OOK) are used for simplicity in the FSO communication. However, in the OOK system it is difficult to choose an optimal detection threshold in presence of the atmospheric turbulence~\cite{peppa10}. In order to enhance the performance of the FSO system, the subcarrier intensity modulation (SIM) is proposed in~\cite{peppa10,popoo09}. The SIM leverages on advances made in signal processing
as well as the maturity of radio frequency (RF) devices such as highly selective filters and stable oscillators, which then
permits the use of modulation techniques such as phase-shift keying (PSK) and quadrature amplitude modulation (QAM)~\cite{peppa10}. 

Cooperative communication is a technique to improve the diversity and coverage area in a distributed antenna based RF communication system~\cite{nosra04,lanem03,sadek10}. %In cooperative communication system, a relaying node cooperates with the source node in mainly two ways: decode-and-forward and amplify-and-forward (AF). 
For FSO systems, cooperative communication has been studied in~\cite{safar07,safar08,karim09}, where the turbulence-induced fading in the FSO links is modeled by the log-normal distribution which is commonly used to model weak turbulence conditions. However, the Gamma-Gamma distribution has been widely adopted for study of behavior of the FSO links under wide range of atmospheric turbulence conditions (weak to strong) because it fits well to the experimental results~\cite{andre01,uysal06,peppa10,popoo09,bayak09}. 

In this letter, we analyze the average bit error rate (BER) performance of the DF based cooperative FSO communication system under the Gamma-Gamma fading channels. We adopt the SIM scheme with the binary phase shift keying (BPSK) modulation. We first derive the CDF and PDF of a random variable (RV) containing mixture of the Gamma-Gamma and Gaussian RVs. Then, with
help of the CDF and PDF expressions, we derive a power series expression of the average
BER of the DF scheme. 
\section{System Model}
Let us consider a \emph{subcarrier intensity-modulated} FSO cooperative system containing a source terminal (S) with two transmit apertures (for transmission of the data to the relay and the destination), one destination node (D) with two receive apertures (for receiving the signals transmitted by the relay and the source), and a single relay (R) with one transmit aperture (for transmitting signals to the destination) and one receive aperture (for receiving signals from the source). 
The transmission from S to D is performed in two phases. In first phase, S transmits BPSK signal $x\in\left\{1,-1\right\}$ to R and D by using different transmit apertures. The relay demodulates the data of S in symbol-wise manner. It is assumed that S uses cyclic redundancy check (CRC) code such that R can know whether it has demodulated the data correctly or not. The relay transmits the demodulated symbols to D in next phase only if it has decoded the data of S correctly.  
The data received in R and D during the first phase will be
\begin{align}
\label{eq:Rdata}
y_{s,d}&=\eta_{s,d} I_{s,d} x+e_{s,d},\nonumber\\
y_{s,r}&=\eta_{s,r} I_{s,r} x+e_{s,r},
\end{align}\\
where $I_{s,r}$ with $E\left\{I_{s,r}\right\}=1$ and $I_{s,d}$ with $E\left\{I_{s,d}\right\}=1$ are the real-valued irradiance of the S-R and S-D links, respectively, following the Gamma-Gamma distribution, $e_{s,r}$ and $e_{s,d}$ are signal-independent complex-valued additive white Gaussian noise (AWGN) with zero mean and variances $N_{s,r}$ and $N_{s,d}$, respectively, and $\eta_{s,d}$ and $\eta_{s,r}$ are the optical-to-electrical conversion coefficients. If R decodes the data of the source correctly, then data received in D in the next phase from R will be \vspace*{-0.75em}%The relay uses an ML decoder and transmits $\hat{x}\in\left\{1,-1\right\}$ (estimated symbol) as follows:
\begin{align}
\label{eq:Ddata}
y_{r,d}=\eta_{r,d} I_{r,d} {x}+e_{r,d},
\end{align}\\
where $I_{r,d}$ with $E\left\{I_{r,d}\right\}=1$ is the real-valued irradiance from R to D, following the Gamma-Gamma distribution, $e_{r,d}$ is the signal-independent complex-valued AWGN noise with zero mean and $N_{r,d}$ variance, and $\eta_{r,d}$ is the optical-to-electrical conversion coefficient. 

The destination uses the following decoder:
\begin{align}
\label{eq:decoder}
U+v V\overset{1}{\underset{-1}{\gtrless}} 0,
\end{align}\\
where $U=4\text{Re}\left\{y^*_{s,d}\eta_{s,d} I_{s,d}/N_{s,d}\right\}$, $V=4\text{Re}\left\{y^*_{r,d}\eta_{r,d} I_{r,d}/N_{r,d}\right\}$, $v=1$ if R transmits, and $v=0$ if R remains silent.\vspace*{-0.75em}
\section{Characterization of A Random Variable containing Mixture of Gamma-Gamma and Gaussian Random Variables}
Let us define a random variable $Y=aZ^2+bZE$, where $a$ and $b>0$ are constants, and $Z$ is Gamma-Gamma distributed as
\begin{align}
\label{eq:ggdist}
f_Z(z)=\frac{2\left(\alpha\beta\right)^{\left(\alpha+\beta\right)/2}}{\Gamma\left(\alpha\right)\Gamma\left(\beta\right)}z^{\left(\alpha+\beta\right)/2-1}K_{\alpha-\beta}\left(2\sqrt{\alpha\beta z}\right),
\end{align}\\
where $K_c\left(\cdot\right)$ is the modified Bessel function of the second kind of order $c$~\cite{abram72}, the parameters $\alpha$ and $\beta$ are related to the atmospheric turbulence conditions~\cite{peppa10,bayak09}, and $E\sim \mathcal{N}\left(0,1\right)$ is the real-valued Gaussian distributed RV. 

Let $W=Z^2$, then the PDF of $W$ can be easily obtained from~\eqref{eq:ggdist} as 
\begin{align}
\label{eq:ggsqaredist}
f_W(w)=\!\frac{\left(\alpha\beta\right)^{\left(\alpha+\beta\right)/2}}{\Gamma\left(\alpha\right)\Gamma\left(\beta\right)}w^{\left(\alpha+\beta\right)\!/4-1}\!K_{\alpha-\beta}\!\!\left(2\sqrt{\alpha\beta}w^{1/4}\right).
\end{align}\\
By using the series expansion of $K_{\alpha-\beta}\left(2\sqrt{\alpha\beta}w^{1/4}\right)$~\cite[Eq.~(9.6.2) and (9.6.10)]{abram72} in~\eqref{eq:ggsqaredist} we have
\begin{align}
\label{eq:ggsqaredistpowser}
f_W(w)\!\!=\!\!\!\!\sum^{\infty}_{k=0}\!\!\left(\!d_k\!\left(\beta,\!\alpha\!\right)\!w^{\beta/2+k/2-1}\!\!\!+\!d_k\!\left(\alpha,\!\beta\right)\!w^{\alpha/2+k/2-1}\!\right),
\end{align}
where 
\begin{align}
\label{eq:dk}
d_k\left(\alpha,\beta\right)\!\!=\!\!\frac{\pi{\left(\alpha\beta\right)}^{\alpha+k}}{2\text{sin}\!\left(\left(\beta-\alpha\right)\pi\right)\!k!\Gamma\!\!\left(\alpha\right)\!\Gamma\!\!\left(\!\beta\right)\Gamma\!\!\left(\alpha\!-\!\beta+k\!+\!1\right)}.
\end{align}\vspace*{0.15em}
\begin{theorem}
The CDF of $Y$ is given as
\begin{align}
\label{eq:cdfy}
F_Y\left(y\right)=\left\{\begin{array}{cc} I\left(\alpha,\beta,-a,b,y\right),&y> 0\\I\left(\alpha,\beta,a,b,0\right),&y= 0\\ I\left(\alpha,\beta,a,b,-y\right),&y<0 \end{array}\right.,
\end{align}\\
where 
\begin{align}
\label{eq:Ifunc1}
I\left(\alpha,\beta,a,b,y\right)\triangleq&\sum^{\infty}_{k=0}\left(d_k\left(\beta,\alpha\right)J\left(a,b,\beta/2+k/2,y\right)\right.\nonumber\\
&\left.+d_k\left(\alpha,\beta\right)J\left(a,b,\alpha/2+k/2,y\right)\right),
\end{align}
\begin{align}
\label{eq:Jfuncsolved1}
J\left(-a,b,m,y\right)=&\sqrt{\frac{1}{2\pi }} \frac{e^{\frac{a y}{b^2}}}{mb} {\frac{y^{m+1/2}}{a^{m-1/2}}} \nonumber\\
&\times \left(K_{m+\frac{1}{2}}\left(\frac{a y}{b^2}\right)+K_{m-\frac{1}{2}}\left(\frac{a y}{b^2}\right)\right), 
\end{align}
\begin{align}
\label{eq:Jfuncsolved2}
J\left(a,b,m,0\right)=\frac{2^{m-1}}{\pi}\left(\frac{b}{a}\right)^{2m}\Gamma\left(m\right) B\left(\frac{1}{2},m+\frac{1}{2}\right),
\end{align}
\begin{align}
\label{eq:Jfuncsolved}
J&\left(a,b,m,-y\right)=\sqrt{\frac{1}{2\pi }} \frac{e^{\frac{a y}{b^2}}}{mb} {\frac{\left(-y\right)^{m+1/2}}{a^{m-1/2}}}\nonumber\\
&\hspace*{4em}\times \left(K_{m+\frac{1}{2}}\left(-\frac{a y}{b^2}\right)-K_{m-\frac{1}{2}}\left(-\frac{a y}{b^2}\right)\right),
\end{align}\newline
and  $B\left(\cdot,\cdot\right)$ is the Beta function~\cite[Eq.~(8.380.1)]{grand07}.
\end{theorem}
\begin{theorem}
The PDF of $Y$ is given as
\begin{align}
\label{eq:pdfy}
f_Y\left(y\right)=\left\{\begin{array}{cc} B\left(\alpha,\beta,-a,b,y\right),&y> 0\\ B\left(\alpha,\beta,a,b,0\right),&y= 0\\B\left(\alpha,\beta,a,b,-y\right),&y<0 \end{array}\right.,
\end{align}
where
\begin{align}
\label{eq:Ifunc}
B&\left(\alpha,\beta,a,b,y\right)\triangleq\sum^{\infty}_{k=0}\left(d_k\left(\beta,\alpha\right)D\left(a,b,\beta/2+k/2-1,y\right)\right.\nonumber\\
&\left.\hspace*{5em}+d_k\left(\alpha,\beta\right)D\left(a,b,\alpha/2+k/2-1,y\right)\right),
\end{align}
\begin{align}
\label{eq:Dfunc1}
D\!\left(-a,b,m,y\right)\!\!=\!\!\sqrt{\frac{2}{\pi b^2}}\!\left(\frac{y}{a}\right)^{m-{1}/{2}}\!\!\!\!\!e^{ay/b^2}\!\!K_{m-1/2}\!\!\left(ay/b^2\!\right),
\end{align}
\begin{align}
\label{eq:Dfunc2}
D\left(a,b,m,0\right)=\frac{\left(2b^2\right)^{m-1}}{\sqrt{\pi}a^{2m-1}}\Gamma\left(m-1/2\right),
\end{align}
%and
\begin{align}
\label{eq:Dfunc3}
D\left(a,b,m,-y\right)=\sqrt{\frac{2}{\pi b^2}}\left(-\frac{y}{a}\right)^{m-{1}/{2}}\!\!\!\!\!e^{ay/b^2}\!\!K_{m-1/2}\!\!\left(-ay/b^2\right)\!.
\end{align}
\end{theorem}
%\begin{proof}
Refer Appendix~\ref{app:1} for proof of Theorem~1 and~2. 
\section{Bit Error Rate of DF Relaying}
Let $x=1$ be the transmitted bit by S but R decides that $-1$ was transmitted, then the probability of error of the S-R link can be written as
\begin{align}
\label{eq:SRerror}
P^{\text{S,R}}_e=\text{Pr}\left(X\leq 0\right),
\end{align}
where $X=4\text{Re}\left\{y^*_{s,r}\eta_{s,r} I_{s,r}/N_{s,r}\right\}$ has the same distribution as that of $Y$ with $a= a_{s,r}=4 \eta^2_{s,r}/N_{s,r}$ and $b=b_{s,r}=2\sqrt{2}\eta_{s,r}/\sqrt{N_{s,r}}$. Therefore, from~\eqref{eq:cdfy} and~\eqref{eq:SRerror}, we have
\begin{align}
\label{eq:SRerror1}
P^{\text{S,R}}_e=I\left(\alpha_{s,r},\beta_{s,r},a_{s,r},b_{s,r},0\right),
\end{align}
where $\left\{\alpha_{s,r},\beta_{s,r}\right\}$ are the turbulence parameters of the S-R link. %, $a_{s,r}=4 \eta^2/N_{s,r}$, and $b_{s,r}=2\sqrt{2}\eta$.
The probability of error in wrongly decoding 1 as -1 in the destination receiver can be written by using~\eqref{eq:decoder} as
\begin{align}
\label{eq:probsubopt}
{P^{\text{D}}_e}&=P^{\text{S,R}}_e\text{Pr}\left(U\leq 0\right)+(1-P^{\text{S,R}}_e)\text{Pr}\left(U+V\leq 0\right),
\end{align}
where the RVs $U$ and $V$ have the same distribution as that of $Y$ with $\left\{a=a_{s,d}=4 \eta^2_{s,d}/N_{s,d},b=b_{s,d}=2 \sqrt{2} \eta_{s,d}/\sqrt{N_{s,d}}\right\}$ and $\left\{a=a_{r,d}=4 \eta^2_{r,d}/N_{r,d},b=b_{r,d}=2 \sqrt{2}\eta_{r,d}/\sqrt{N_{r,d}}\right\}$, respectively. We can rewrite~\eqref{eq:probsubopt} as
\begin{align}
\label{eq:probsubopt1}
{P^{\text{D}}_e}&=P^{\text{S,R}}_e P^{\text{S,D}}_e+(1-P^{\text{S,R}}_e)\text{Pr}\left(U+V\leq 0\right),
\end{align}
where $P^{\text{S,D}}_e$ can be obtained similar to~\eqref{eq:SRerror1} and it is shown in Appendix~\ref{app:2} that for $a_{s,d}b^2_{r,d}=a_{r,d}b^2_{s,d}$, we have
\begin{align}
\label{eq:int}
&\text{Pr}\left(U+V\leq 0\right)=\sum^\infty_{k_1=0}\sum^\infty_{k_2=0}
\displaystyle\left\{d_{k_1}\left(\beta_{r,d},\alpha_{r,d}\right)d_{k_2}\left(\beta_{s,d},\alpha_{s,d}\right)\right.\nonumber\\
&\left. \times M\left(\bs{g},\beta_{r,d}/2+k_1/2,\beta_{s,d}/2+k_2/2\right)+d_{k_1}\left(\alpha_{r,d},\beta_{r,d}\right)\right.\nonumber\\
&\left.\times d_{k_2}\left(\beta_{s,d},\alpha_{s,d}\right) M\left(\bs{g},\alpha_{r,d}/2+k_1/2,\beta_{s,d}/2+k_2/2\right)\right.\nonumber\\
&\left. +d_{k_1}\left(\alpha_{r,d},\beta_{r,d}\right)d_{k_2}\left(\alpha_{s,d},\beta_{s,d}\right)\right.\nonumber\\
&\left. \times M\left(\bs{g},\alpha_{r,d}/2+k_1/2,\alpha_{s,d}/2+k_2/2\right)+d_{k_1}\left(\beta_{r,d},\alpha_{r,d}\right)\right.\nonumber\\
&\left. \times d_{k_2}\!\!\left(\alpha_{s,d},\!\beta_{s,d}\right)\!M\!\left(\bs{g},\beta_{r,d}/2\!+\!k_1/2,\alpha_{s,d}/2\!+\!k_2/2\right)\right\},
\end{align}
where $\bs{g}=\left[a_{r,d},b_{r,d},a_{s,d},b_{s,d}\right]$, and $\left\{\alpha_{r,d},\beta_{r,d}\right\}$ and $\left\{\alpha_{s,d},\beta_{s,d}\right\}$ are the turbulence parameters of the R-D and S-D links, respectively;
\begin{align}
\label{eq:int2}
&M\left(\tilde{\bs{g}},m_1,m_2\right)=\frac{2^{m_1+m_2-1}}{\sqrt{\pi}m_2b^{2m_1}_1}\frac{b^{4m_1+2m_2}_2}{a^{2m_1+2m_2}_2}\nonumber\\
&\hspace*{4em}\times\frac{\Gamma\left(m_1+m_2+1/2\right)\Gamma\left(m_1\right)\Gamma\left(m_2+1\right)}{\Gamma\left(m_1+m_2+1\right)}\nonumber\\
&\times _2\!\!F_1\!\!\left(\!\!m_1\!+\!m_2\!+\!\frac{1}{2},m_2\!+\!1;m_1\!+\!m_2+1;1\!-\!\frac{a^2_1b^4_2}{b^4_1a^2_2}\right),
\end{align}\\
where $\tilde{\bs{g}}=\left[a_1,b_1,a_2,b_2\right]$, $_2F_1\left(\cdot,\cdot;\cdot;\cdot\right)$ is the Hypergeometric function~\cite{grand07}, and $d_{k_1}\left(\alpha_{r,d},\beta_{r,d}\right)$ and $d_{k_2}\left(\alpha_{s,d},\beta_{s,d}\right)$ are given in~\eqref{eq:dk}.
\vspace*{-1em}
\section{Numerical Results}
The analysis and simulation is done under strong ($\alpha=4.2$, $\beta=1.4$) and moderate ($\alpha=4.0$, $\beta=1.9$) atmospheric turbulence conditions~\cite{popoo09}. It is assumed that S-R, S-D, and R-D link have the same average signal-to-noise ratio (SNR) and atmospheric turbulence conditions and $\eta_{s,d}=\eta_{s,r}=\eta_{r,d}=1$.  

\begin{figure}[t!]\vspace*{-1.0em}
  \begin{center}\hspace*{-1.8em}
    \psfig{figure=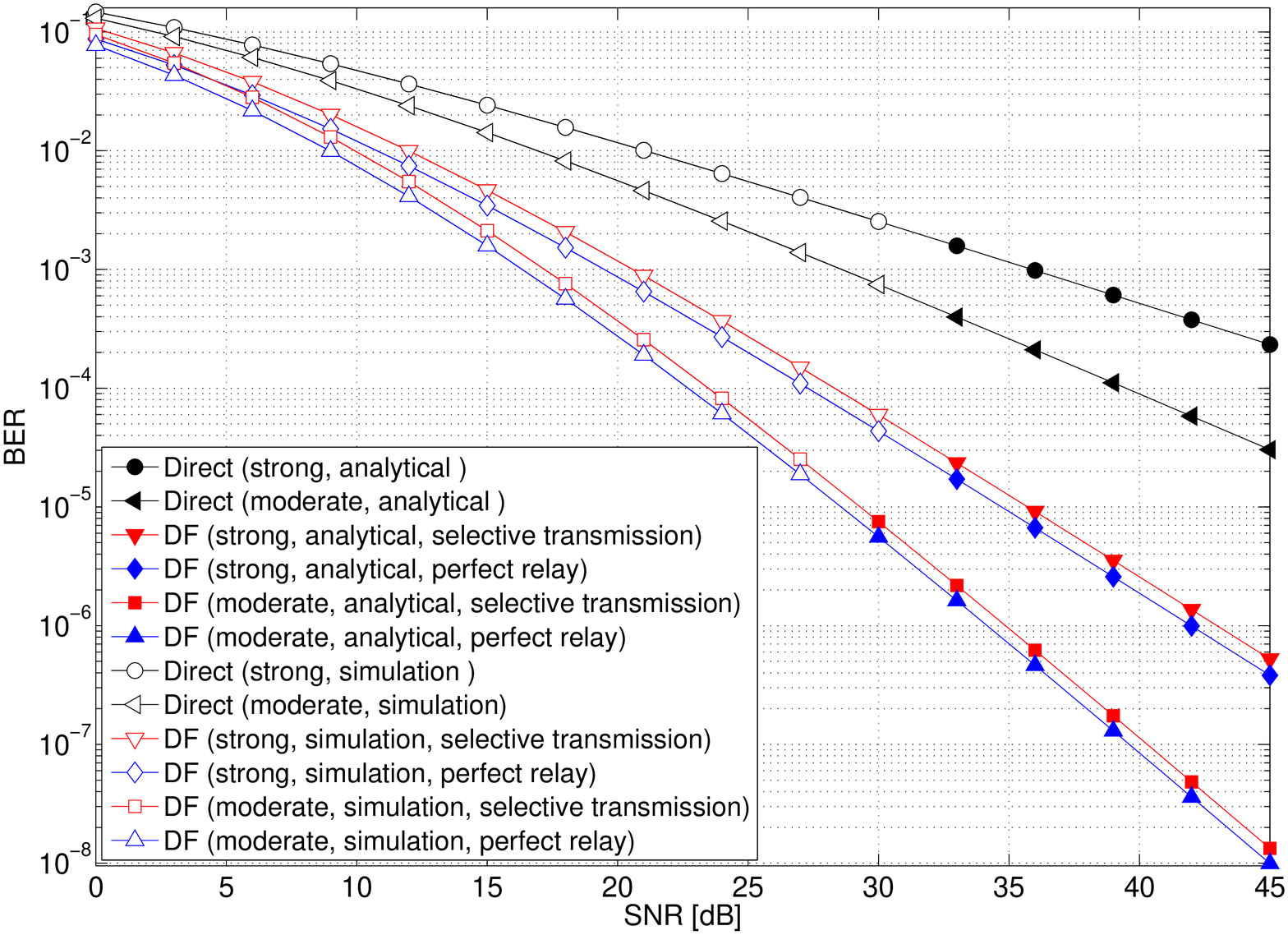,height=2.60in,width=3.9in}
    \vspace*{-1.5em}
    \mycaption{Average BER of the DF relaying with SIM BPSK in Gamma-Gamma fading channels.}
    \label{fig:comp_ML_UI}
    \vspace*{-1.75em}
  \end{center}
\end{figure}

In Fig.~\ref{fig:comp_ML_UI}, we have plotted BER versus SNR performance of the three node cooperative FSO system employing SIM with BPSK and using DF protocol with selective relay which transmits only if it decodes the data of S correctly and perfect relay which commits no error in detection, in strong and moderate Gamma-Gamma fading channels which vary independently in every time-interval. The SNR on y-axis of  Fig.~\ref{fig:comp_ML_UI} indicates the average SNR of the S-D link. We have also plotted BER versus SNR performance of a non-cooperative direct transmission based system, where S transmits at double power than the cooperative FSO system. It can be seen from Fig.\!~\ref{fig:comp_ML_UI} that the cooperative FSO system significantly outperforms the direct transmission based system under strong and moderate turbulence conditions. For example, the selective relay based DF FSO cooperative system can achieve an SNR gain of 14~dB over the non-cooperative scheme for a moderate turbulence regime at BER=$10^{-4}$. Moreover, the selective transmission based DF scheme have much better diversity than the direct transmission scheme and works very close to the DF scheme with perfect relay under strong and moderate turbulence conditions. 
\vspace*{-1.0em}
\section{Conclusions}
We have analyzed the DF based FSO cooperative system over the Gamma-Gamma fading channels. The advantages of using cooperative communication over a direct transmission based non-cooperative system are established by using analysis and simulation. It is shown that the BER of the DF FSO system significantly reduces at high SNR as compared to the non-cooperative system. \vspace*{-1em}
\appendices
\section{Derivation of CDF and PDF of $Y$}
\label{app:1}
Since $f_{Y|Z,a,b}(y)\sim\mathcal{N}\left(aW,b^2W\right)$, the CDF of $Y$, i.e., $F_Y\left(y\right)=\text{Pr}\left(Y\leq y\right)={E}_{W}\left\{1-\text{Pr}\left(Y\geq y|W,a,b\right)\right\}$, can be written by using~\cite[Eqs.~(2.3.12) and~(2.3.14)]{proak08} as
\begin{align}
\label{eq:cdfq}
F_Y\left(y\right)=\left\{\begin{array}{cc}1-\int^\infty_0Q\left(\frac{y-aw}{\sqrt{b^2w}}\right)\! f_W\left(w\right) dw,& y\geq 0\\
\int^\infty_0Q\left(\frac{aw-y}{\sqrt{b^2 w}}\right) f_W\left(w\right) dw,& y< 0
  \end{array}\!,\right.
\end{align}\\
where $Q\left(\cdot\right)$ is the q-function~\cite[Eq.~(2.3.10)]{proak08}.
From~\eqref{eq:ggsqaredistpowser},~\eqref{eq:dk}, and~\eqref{eq:cdfq}, we get~\eqref{eq:cdfy} in terms of $I\left(\alpha,\beta,a,b,y\right)$ defined in~\eqref{eq:Ifunc1}, where % and $J\left(a,b,m,y\right)$ defined as
\begin{align}
\label{eq:Jfunc}
J\left(a,b,m,y\right)=\int^\infty_0w^{m-1}Q\left(\frac{y+aw}{\sqrt{b^2 w}}\right)dw, \ms m\geq 0.
\end{align}\\
By using the following relation:
$Q\left(x\right)=\frac{1}{2}\text{erfc}\left(\frac{x}{\sqrt{2}}\right),$
where $\text{erfc}\left(\cdot\right)$ is the complimentary error function,
and\cite[Eq.~(2.8.4.3)]{prudn92} in~\eqref{eq:Jfunc}, 
we get~\eqref{eq:Jfuncsolved1} and~\eqref{eq:Jfuncsolved}.  
By subtituting $y=0$
and using the following relation: 
$Q\left(x\right)=\frac{1}{\pi}\int^{\pi/2}_0e^{-\frac{x^2}{2\text{sin}^2\theta}}d\theta, \ms x\geq 0,
$~\cite[Eq.~(3.381.4)]{grand07}, and~\cite[Eq.~(3.191.3)]{grand07} in~\eqref{eq:Jfunc}, we get~\eqref{eq:Jfuncsolved2}. 

By taking partial derivative of~\eqref{eq:cdfy} with respect to $y$, we get~\eqref{eq:pdfy} in terms of $B\left(\alpha,\beta,a,b,y\right)$ defined in~\eqref{eq:Ifunc}, where %Let us define
\begin{align}
\label{eq:Ddef}
\hspace*{0.75em}D\left(a,b,m,y\right)= \frac{\delta}{\delta y}J\left(a,b,m,y\right).
\end{align}\\
{The right hand side of~\eqref{eq:Ddef} can be simplified by using~\eqref{eq:Jfunc},~\cite[Eq.~(0.410)]{grand07}, and the relation $\frac{d}{d x}Q(x)
=-\frac{1}{\sqrt{2\pi}}e^{-x^2/2}$.  
Then we can obtain~\eqref{eq:Dfunc1} and~\eqref{eq:Dfunc3} by using ~\cite[Eq.~(3.471.9)]{grand07}, and~\eqref{eq:Dfunc2} can be obtained by substituting $y=0$ and using~\cite[Eq.~(3.381.4)]{grand07}.}\vspace{-1.0em} 
\section{Derivation of $\text{Pr}\left(U+V\leq 0\right)$}
\label{app:2}
After some algebra, we can write
\begin{align}
\label{eq:PrUV}
&\text{Pr}\left(U+V\leq 0\right)\!\!
=\!\!\!\int^\infty_{0}\!\!\!\!\!\!\!F_U\!\!\left(v\right)\!\!f_V\!\!\left(-v\right)\!\!\:dv\!+\!\!\! \int^\infty_{0}\!\!\!\!\!\!\!F_U\!\!\left(-v\right)\!\!f_V\!\!\left(v\right)\!\!\:dv.
\end{align}\\
Let us define the following function:
\begin{align}
\label{eq:M}
&M\left(\tilde{\bs{g}},m_1,m_2\right)\triangleq\frac{2}{\pi b_1b_2m_2a^{m_1-1/2}_1a^{m_2-1/2}_2}\nonumber\\
&\times\!\!\!\int^\infty_{0}\!\!\!\!\!\!\!y^{m_1+m_2}K_{m_1-1/2}\left(a_1y/b^2_1\right) K_{m_2+1/2}\left(a_2y/b^2_2\right)dy.
\end{align}\\
Equation~\eqref{eq:int} can be obtained by using~\eqref{eq:cdfy},~\eqref{eq:pdfy},~\eqref{eq:PrUV},~\eqref{eq:M}, and~\cite[Eq.~(2.16.33.1)]{prudn92}.\vspace*{-0.75em}
\bibliography{IEEEabrv,biblitt}
\bibliographystyle{IEEEtran}
\end{document}